\documentclass{appolb}
\usepackage{graphicx,wrapfig}
\begin{document}
\title{\mbox{No serious meson spectroscopy without
scattering\thanks{Talk by G.~Rupp at Workshop ``EEF70'', Coimbra,
Portugal, September 1--5, 2014.}}
}
\author{George Rupp\address{CFIF, Instituto Superior T\'{e}cnico,
Universidade de Lisboa, P-1049-001, Portugal}
\\[5mm]
Eef van Beveren\address{CFC, Departamento de F\'{\i}sica,
Universidade de Coimbra, P-3004-516, Portugal}
\\[5mm]
Susana Coito\address{Institute of Modern Physics, CAS, Lanzhou 730000, China}
}
\maketitle

\begin{abstract}
The principal purpose of meson spectroscopy is to understand the confining
force, which is generally assumed to be based on low-energy QCD. This is
usually done in the context of quark models that ignore the dynamical
effects of quark-pair creation and decay. Very recent lattice calculations
confirm much earlier model results showing that neglecting such effects,
in the so-called quenched approximation, may give rise to discrepancies of
hundreds of MeV, and so distort the meson spectra resulting from quark
confinement only. Models attempting to mimic unquenching through a
redefinition of the constituent quark mass or screening of the confining
potential at larger interquark separations are clearly incapable of
accounting for the highly non-perturbative and non-linear effects on mesonic
bound-state and resonance poles, as demonstrated with several published
examples.
\end{abstract}

\section{Introduction}

The experimentally observed spectra of mesons and baryons should provide
detailed information on quark confinement and other interquark forces, which
are believed to result from QCD at low energies. Thus, over the past four
decades work on quark models has attempted to reproduce these spectra by
employing confining potentials, which are usually of a Coulomb-plus-linear or
``funnel'' type, on the basis of short-distance perturbative QCD and
long-distance QCD speculations. A typical and often cited example is the
relativised quark model of Godfrey and Isgur \cite{PRD32p189}. In such
approaches, it is generally assumed that hadronic decay can be treated \em a
posteriori \em \/and perturbatively, with no appreciable influence on the
spectrum itself.

However, very recent unquenched lattice calculations have shown the sizable
effects of accouting for dynamical quarks and allowing hadrons to decay
strongly. In particular, in Ref.~\cite{PRD88p054508} a lattice computation
of $P$-wave $K\pi$ phase shifts and the lowest strange vector-meson resonances
was carried out, confirming $K^\ast(892)$ and tentatively also
$K^\ast(1410)$, though the latter resonance came out about 80~MeV below the
experimental \cite{PDG2014} mass. More surprisingly, an equally unquenched
calculation by the same lattice collaboration \cite{ARXIV13116579}, yet without
two-meson correlators and so no hadronic decay, predicted a (bound)
$K^\ast(1410)$ state roughly 300~MeV higher in mass. Despite possible
inaccuracies due to problems in dealing with inelastic resonances and very
light quarks/pions on the lattice, this enormous difference confirms the
importance of including strong decay for reliable predictions in meson
spectroscopy. Moreover, in another study by still the same lattice group, the
low mass of the charmed-strange $D_{s0}^\ast(2317)$ meson was reproduced by
including two-meson correlators corresponding to the subthreshold $S$-wave
$DK$ channel \cite{PRL111p222001}, in agreement with an unquenched quark-model
description a decade earlier \cite{PRL91p012003} (also see below).

As already said, lattice QCD still faces considerable problems in
dealing with resonances that have multiple decay modes and in extrapolating
predictions towards the physical pion mass, besides serious difficulties in
dealing with heavy quarkonia and excited states. Therefore, in the foreseeable
future QCD-inspired quark models will still be of a crucial importance in
interpreting and advising on experiments in hadron spectroscopy. Clearly, such
models should go beyond the quenched approximation of the confinement-only
approaches mentioned above.

One attempt to do this in a ``cheap'' way amounted
to estimating hadron-loop mass corrections in charmonium \cite{PRC77p055206},
and suggesting that to a large extent it might suffice to adjust the charm
quark mass. However, this is most likely a too simplistic assessment, since the
size of hadronic loop corrections will depend on the wave function of a
specific state, in particular its nodal structure, in view of the peaked shape
of the string-breaking interaction leading to decay, as confirmed on the
lattice \cite{PRD71p114513}. Furthermore, above the lowest decay threshold the
effects will be governed by $S$-matrix unitarity and analyticity, which are
generally non-perturbative and non-linear (see the examples below), except for
unphysically small couplings \cite{PTP125p581}.

Another approach to mimicking unquenching in quark models is by screening the
confining potential at larger interquark separations, making it in fact
non-confining and so allowing for decay. However, such decays are pathological,
as they lead to free quarks and not hadrons in the final state. Of course, one
can adjust the model parameters such that the thresholds for decay into free
quarks lie above all experimentally observed states. However, then one would
purport to describe physical resonances by treating them as stable hadrons,
ignoring effects due to $S$-matrix analyticity and genuine decay thresholds.
Moreover, the usually employed screened potentials in quark models are,
for short distances, similar to the funnel potential \cite{PRD32p189}, as 
e.g.\ in the model of Ref.~\cite{PRD78p114033}, which on top of that has
interquark meson exchanges. Thus, these models share some of the shortcomings
\cite{APPS5p1007} of the Godfrey-Isgur model \cite{PRD32p189}, such as much too
large radial separations for the lowest states in the light-quark sector.

In this talk, we shall make the case for an $S$-matrix approach to meson
spectroscopy, as the only reliable phenomenological way to unquench the quark
model, describing both mesonic bound states and resonances in a unique
analytic formalism, with meson-loop effects included to all orders. For that
purpose published work on several enigmatic meson resonances will be briefly
reviewed. The organisation is as follows: Sec.~2 deals with the original model
and the vector charmonium spectrum, Sec.~3 with the light scalar mesons. Sec.~4
with the charmed scalars $D_{s0}^\ast(2317)$ and $D_0^\ast(2300)$, Sec.~5 with
the axial-vector charmonium state $X(3872)$, and Sec.~6 with the $J^P=1^+$
open-charm mesons $D_{s1}(2536)$, $D_{s1}(2460)$, $D_1(2420)$, and $D_1(2430)$.
A few conclusions are presented in Sec.~7.

\section{HO model and vector charmonium spectrum}
\begin{wrapfigure}{r}{3cm}
\vspace*{-3mm}
\includegraphics[trim = 60mm 53mm 70mm 51mm,clip,width=3cm,angle=0]
{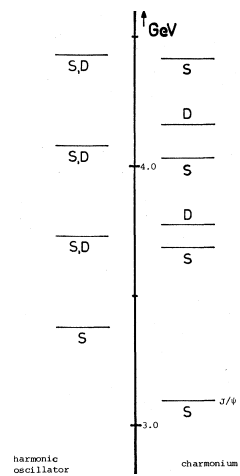}
\caption{HO vs.\ vector $c\bar{c}$ states \cite{PRD21p772}.}
\label{HOvector}
\end{wrapfigure}
The first three vector charmonium levels, discovered almost four decades ago,
seem to suggest a confining potential with decreasing splittings for 
increasing radial quantum number. Such a pattern can result from a power-law
potential $r^n$ with $n<2$. The simplest case is a linear potential, but also
a funnel-type potential, which includes a Coulombic piece, will do, as e.g.\ in
the model by Godfrey and Isgur \cite{PRD32p189}. However, the coupled-channel
model of Ref.~\cite{PRD21p772}, employing an HO potential with constant
frequency and a transition potential mimicking string breaking at a certain
distance to describe OZI-allowed strong decay, leads to a similar result.
Figure~\ref{HOvector} shows schematically how the equidistant HO spectrum, with
degenerate $S$ and $D$ states, is transformed into the physical charmonium
spectrum by unquenching. Moreover, even the first few bottomomium states are
automatically reproduced \cite{PRD21p772}.
\section{$f_0(500)$ and other light scalar mesons}
\begin{wrapfigure}{r}{6cm}
\vspace*{-4mm}
\includegraphics[trim = 30mm 170mm 45mm 45mm,clip,width=6cm,angle=0]
{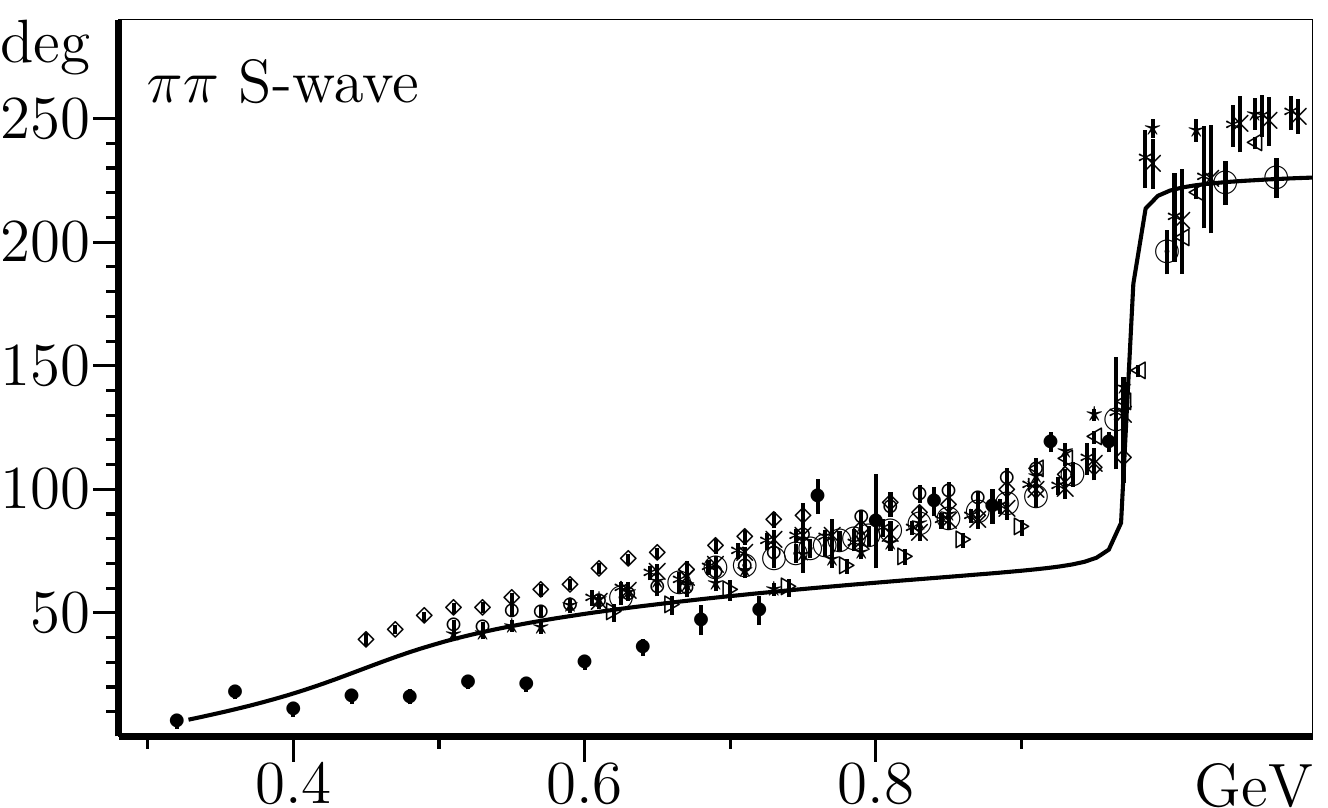}
\caption{$S$-wave $\pi\pi$ phase shifts \cite{ZPC30p615}.}
\label{pipiS}
\end{wrapfigure}
An extended version of the above unquenched HO model
\cite{PRD21p772}, with a smeared-out transition potential and relativistic
kinematics in the open decay channels, was applied to heavy and light vector
and pseudoscalar mesons \cite{PRD27p1527}. This very same model was then
used, with unchanged parameters, to study scalar mesons made of light quarks
\cite{ZPC30p615}. The resulting $S$-matrices revealed resonance poles not
only in the expected 1.3--1.5 GeV energy region, but also well below 1 GeV.
It comprised a complete extra scalar nonet, including the then still very
controversial $f_0(500)$ (``$\sigma$'') and $K_0^\ast(800)$ (``$\kappa$'')
mesons, with pole positions \cite{ZPC30p615} close to present-day world
averages \cite{PDG2014}. In Fig.~\ref{pipiS} the model's parameter-free
prediction of $S$-wave $\pi\pi$ phase shifts is shown with old data.

\section{Charmed scalars $D_{s0}^\ast(2317)$ and $D_0^\ast(2300)$}
\begin{figure}[h!]
\vspace*{-7mm}
\hfill
\includegraphics[trim = 50mm 94mm 40mm 47mm,clip,width=4cm,angle=0]
{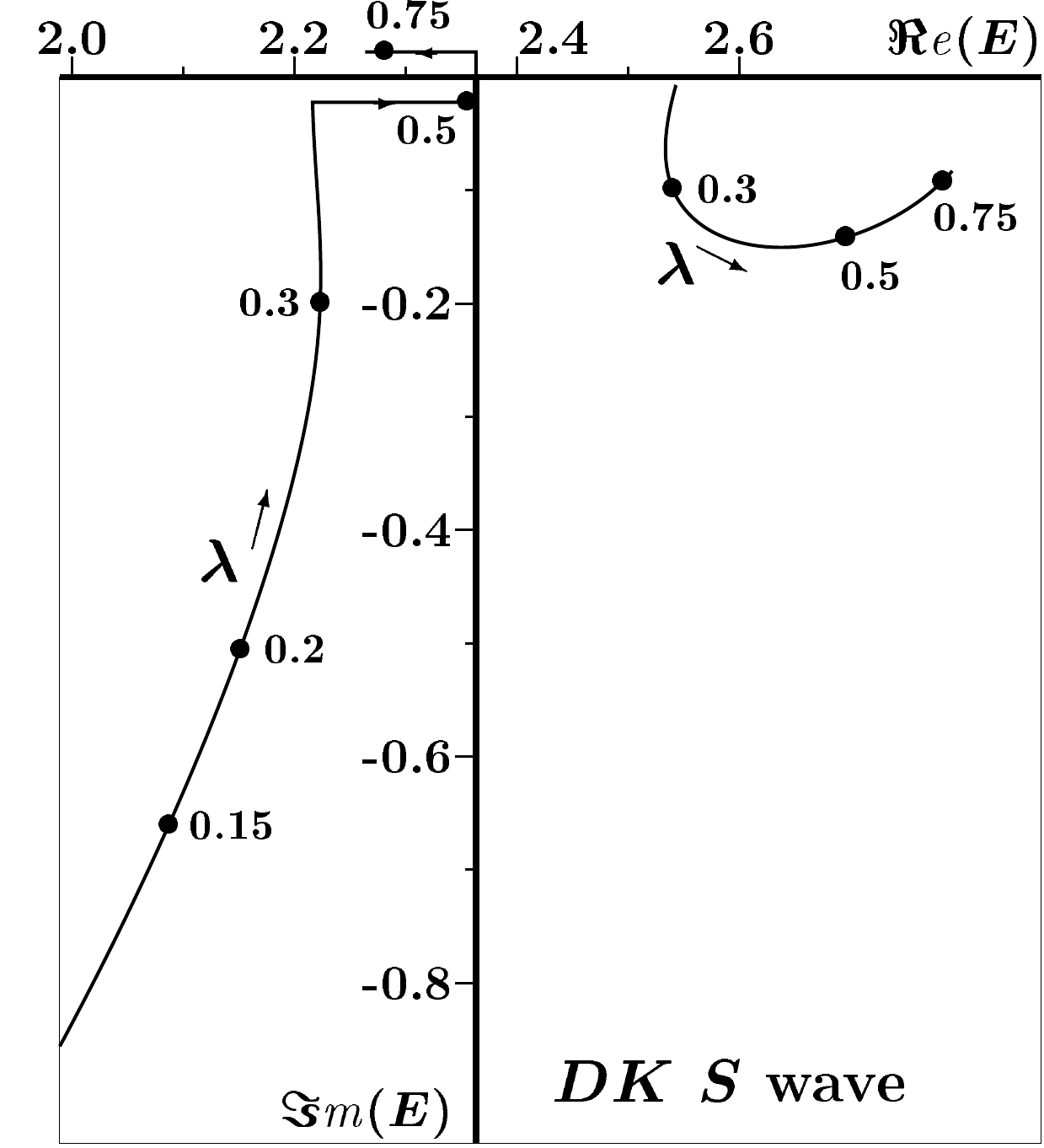}
\hfill
\includegraphics[trim = 45mm 93.7mm 30mm 40mm,clip,width=4.45cm,angle=0]
{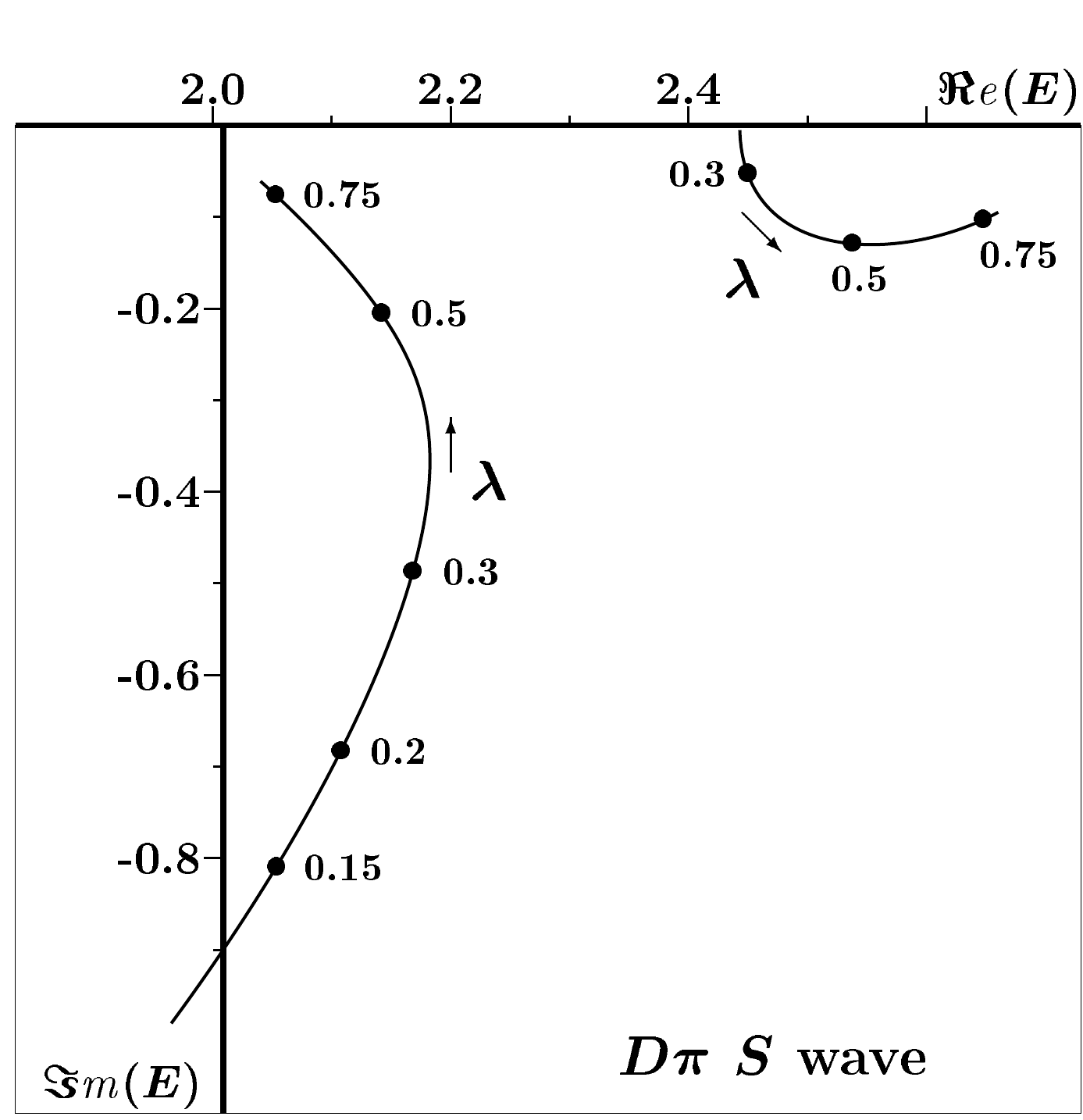}
\hfill
\mbox{ }
%\end{tabular}
\caption{Pole trajectories of $D_{s0}^\ast(2317)$ (left) and $D_0^\ast(2300)$
(right) as a fuction of $\lambda$ \cite{PRL91p012003}, with $\lambda=0.75$ the
physical value. Higher recurrences are also shown.}
\label{charmscalars}
\end{figure}

A momentum-space version \cite{AOP324p1620} of the above model
\cite{PRD27p1527} was applied \cite{PRL91p012003} to the $D_{s0}^\ast(2317)$
and $D_0^\ast(2300)$ \cite{PDG2014} charmed scalar mesons, in a very simple
approximation for the lowest states, but with quark masses and HO frequency 
fixed at the values determined in Ref.~\cite{PRD27p1527}. This very same model,
with identical values for the overall coupling $\lambda$ and the
string-breaking distance $r_0$, had been used before in an excellent fit
\cite{EPJC22p493} to the $S$-wave $K\pi$ phase shifts up to 1.5 GeV, while
simultaneusly reproducing the $K_0^\ast(800)$ and $K_0^\ast(1430)$
 \cite{PDG2014} resonances. The resulting \cite{PRL91p012003} pole trajectories
as a function of $\lambda$ are shown in Fig.~\ref{charmscalars}, with an
excellent prediction for $D_{s0}^\ast(2317)$.

\section{Axial-vector charmonium state $X(3872)$}
\begin{wrapfigure}{r}{5cm}
\vspace*{-3mm}
\includegraphics[trim = 25mm 121mm 50mm 6mm,clip,width=5cm,angle=0]
{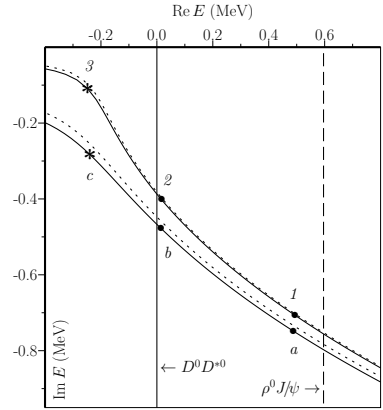}
\caption{$X(3872)$ pole trajectories for different parameters
(see Ref.\ \cite{EPJC71p1762} for details).}
%right: $c\bar{c}$ and $D^0D^{*0}$
%wave-function components (see Ref.~\cite{EPJC73p2351} for details).}
\label{x3872}
\end{wrapfigure}
The $J^{PC}=1^{++}$ charmonium state $X(3872)$ \cite{PDG2014} is a perfect
laboratory for quark and effective models, as it is bound by only 0.11~MeV with
respect to its lowest OZI-allowed decay channel, i.e., $D^0D^{\ast0}$. This
system was studied recently in a multichannel momentum-space model and also in
a two-component coordinate-space model. The former \cite{EPJC71p1762}
demonstrated that an $X(3872)$ resonance pole with a realistic imaginary part
(see Fig.~\ref{x3872}) can result from unquenching a bare $2\,{}^{3\!}P_1$
$c\bar{c}$ state about 100 MeV higher in mass via several two-meson channels,
including the OZI-forbidden $\rho^0 J\!/\!\psi$ and $\omega J\!/\!\psi$
channels, and accounting for the $\rho^0$, $\omega$ widths. On the other hand,
the latter paper \cite{EPJC73p2351} showed that $X(3872)$ has a sizable
$c\bar{c}$ component and thus cannot be considered a $D^0D^{\ast0}$ molecule
(also see Ref.~\cite{ARXIV14111654} and talk by M.~Cardoso
\cite{ARXIV14127406}).

\section{\mbox{$1^+$ charmed mesons $D_{s1}(2536)$, $D_{s1}(2460)$,
$D_1(2420)$, $D_1(2430)$}}
The charmed-strange and charmed-light mesons with $J^P=1^+$ reveal
\cite{PDG2014} a very irregular pattern of masses and widths, impossible to
understand with the ususal perturbative spin-orbit interactions and decay
amplitudes, or heavy-quark effective theory. However, a multichannel
\begin{figure}[bh]
\begin{tabular}{cc}
\hspace*{1cm}
\includegraphics[trim = 38mm 121mm 46mm 24mm,clip,height=5cm,angle=0]
{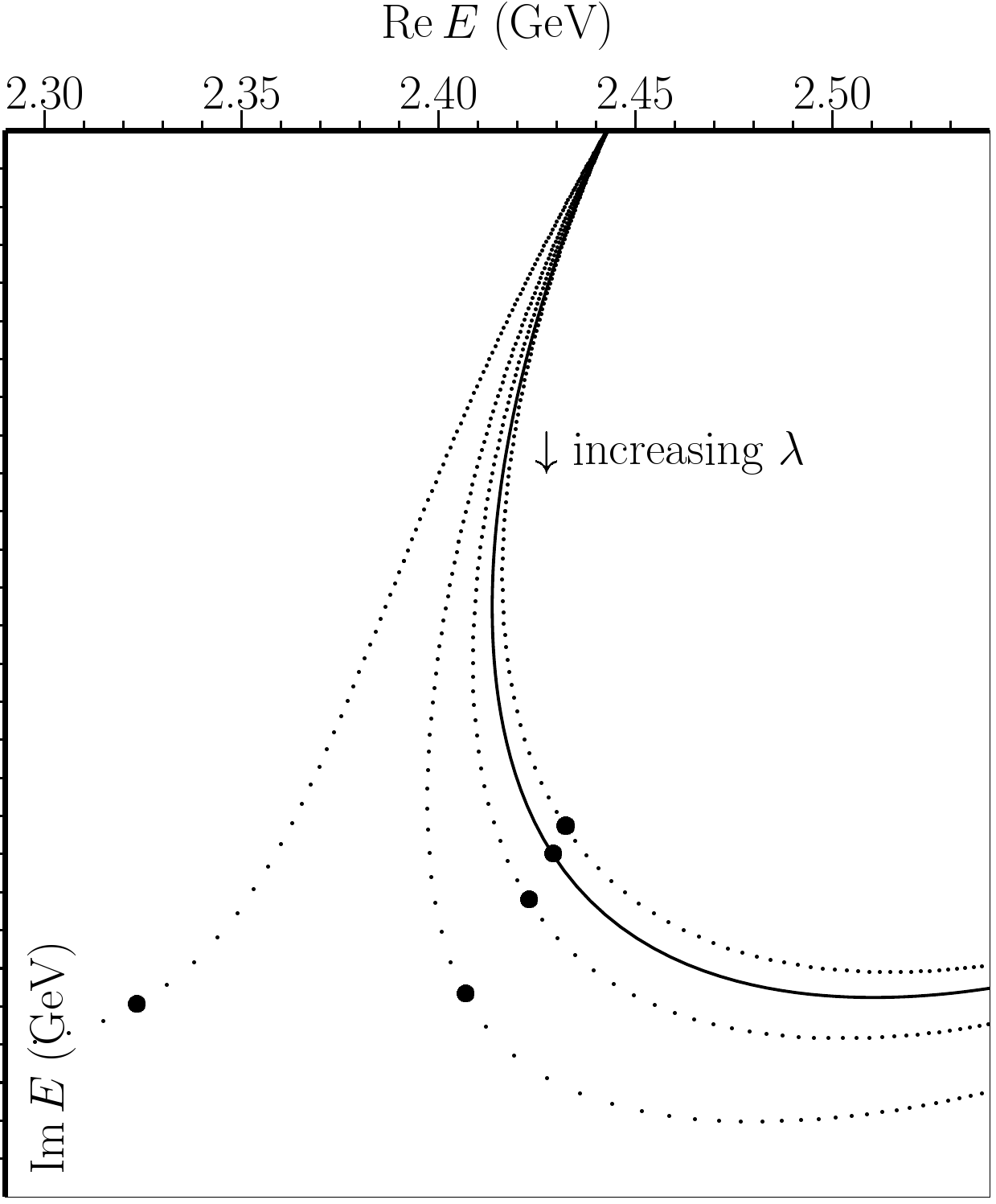}
&
\hspace*{1cm}
\includegraphics[trim = 21mm 121mm 51mm 6mm,clip,height=5cm,angle=0]
{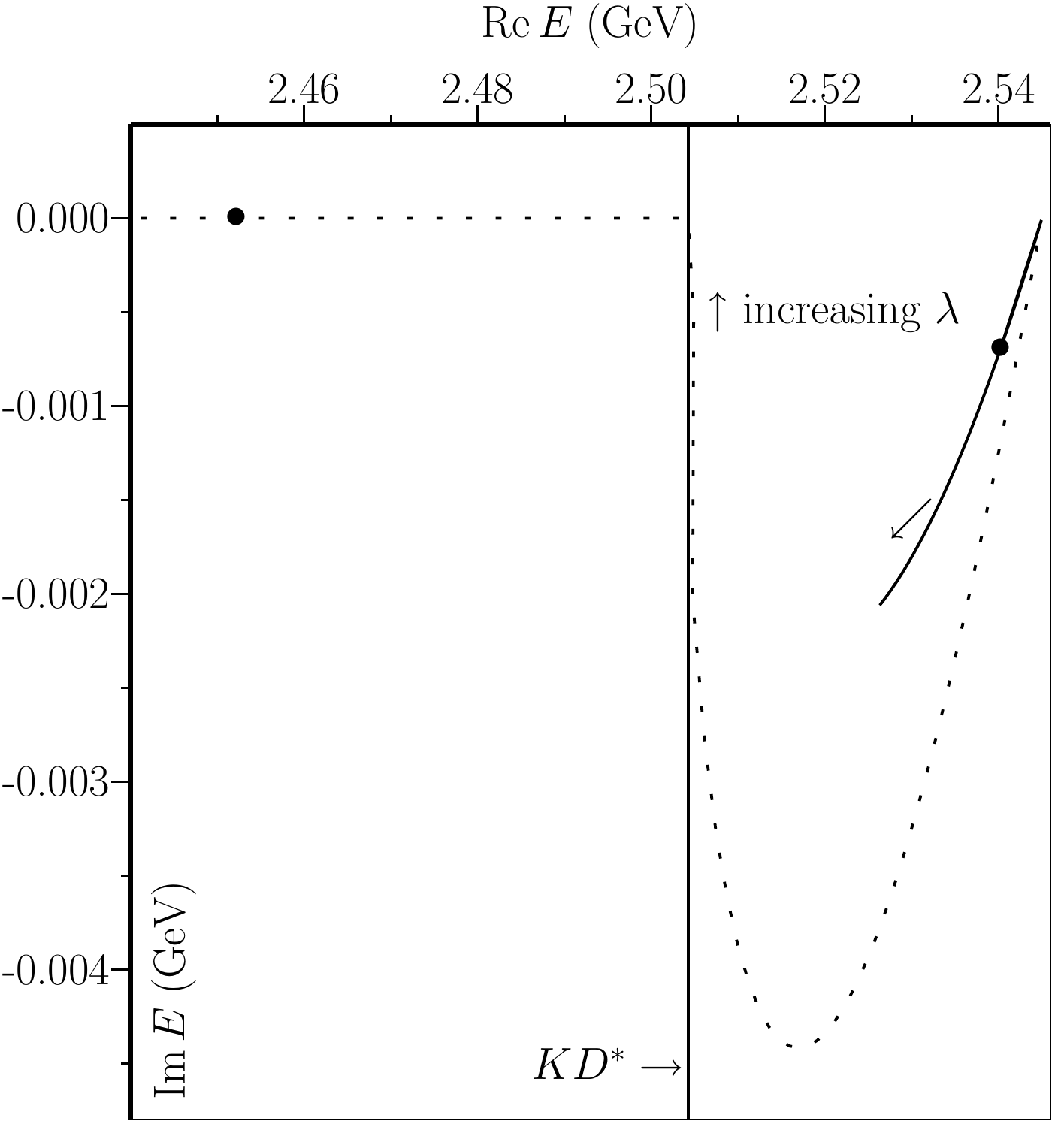}
\end{tabular}
\caption{Pole trajectories of $D_1(2430)$ (left), $D_{s1}(2460)$ (right,
dotted) and $D_{s1}(2536)$ (right, solid). For further details,
see Ref.~\cite{PRD84p094020}.}
\label{axialcharm}
\end{figure}
unquenched quark-model calculation with bare $1\,{}^{3\!}P_1$ and
$1\,{}^{1\!}P_1$ $c\bar{s}$ or $c\bar{q}$ seeds does an
excellent job in reproducing the data \cite{PRD84p094020}. In
Fig.~\ref{axialcharm} some pole trajectories are shown.

\section{Conclusions}
Meson spectroscopy is truly different from atomic spectroscopy, in that line
widths can be of the same order as level spacings. $S$-matrix analyticity then
implies that real level shifts may be of similar or even greater magnitude.
The proper, non-perturbative way to deal with this is by describing mesons as
resonances or bound states in a scattering process of the dominant real or 
virtual decay products, yet while dealing with quark confinement at the same
time and on an equal footing. The above examples from various sectors of the
meson spectrum should provide support for such an approach. \\
In conclusion, meson spectroscopy is even more involved because of natural,
non-resonant threshold enhancements (see talk by E.~van Beveren
\cite{ARXIV150104862}).

\end{document}